\documentclass[12pt]{article}
\usepackage{mathtools,amssymb,mathrsfs,ytableau,microtype,tikz,slashed}
\usepackage{physics}
\usepackage{color}
\usepackage[linktocpage]{hyperref}
\hypersetup{colorlinks=true,citecolor=red,linkcolor=blue, urlcolor=blue, breaklinks=true}
\usepackage{cite}
\usepackage{moresize}
\usepackage{url}
\ytableausetup{centertableaux,boxsize=.5em}
\usepackage[numbers,sort]{natbib}

\usepackage{pifont}

\makeatletter\@addtoreset{equation}{section}\makeatother

\renewcommand{\title}[1]{\vbox{\center\LARGE{#1}}\vspace{5mm}}
\renewcommand{\author}[1]{\vbox{\center\large#1}\vspace{5mm}}
\newcommand{\address}[1]{\vbox{\center\em#1}}

\definecolor{ao}{rgb}{0.13, 0.55, 0.13}

\newcommand{\cmark}{{\color{ao}\ding{51}}} 
\newcommand{\xmark}{{\color{red}\ding{55}}}

\begin{document}

\begin{titlepage}

\begin{center}

\vskip 0.5cm

\title{
Detecting Standard Model Gauge Group from \\[+5pt] 
Generalized Fractional Quantum Hall Effect
}

 \author{Po-Shen Hsin$^{1}$, Jaume Gomis$^{2}$
 }

\vskip 0.5cm

\address{${}^1$ Department of Mathematics, King’s College London, \\
Strand, London WC2R 2LS, UK.}

\vskip 0.5cm
\address{${}^2$ Perimeter Institute for Theoretical Physics,\\
Waterloo, Ontario, N2L 2Y5, Canada.}

\vskip 1cm

\end{center}

\abstract{\normalsize{The Standard Model of particle physics stands as one of the most profound and successful frameworks for describing the fundamental workings of nature. The global form  of the Standard Model gauge group, however,  remains an open question: it can be $\left(SU(3)_C\times SU(2)_W\times U(1)_Y\right)/\Gamma$ with  $\Gamma=1,\mathbb{Z}_2,\mathbb{Z}_3$ or $\mathbb{Z}_6$. The work introduces      the  fractional topological transport coefficient  $\xi$  involving
the  $U(1)$  B-L   symmetry and the $U(1)$ one-form magnetic symmetry  of the renormalizable Standard Model, and show that it distinguishes the global form of the Standard Model gauge group. The gauge group is  fully determined   for specific  values of $\xi$, which also depends on the choice of action of the  B-L symmetry on the Standard Model  known as a fractionalization class. This transport coefficient can be measured in    a contact term for the two-point function of the B-L symmetry current with the magnetic one-form symmetry current of the Standard Model.  This   parallels   topological transport in the Fractional Quantum Hall Effect, with quarks and the B-L symmetry playing the role of anyons and   the $U(1)$ electromagnetic global symmetry respectively.

}}

\vfill

\today

\vfill

\end{titlepage}

\eject

\tableofcontents

\unitlength = .8mm

\setcounter{tocdepth}{3}

\section{Introduction, Summary and  Conclusion}

The Standard Model of particle physics stands as a triumph of theoretical physics, providing an unparalleled description of a vast array of experimental data with remarkable precision. In spite of its many successes,      several fundamental questions remain  unanswered.

While the $su(3)\times su(2)\times u(1)$  gauge algebra of the Standard Model, along with the quantum numbers of quarks, leptons and the Higgs boson, are experimentally determined, the exact global structure of the Standard Model gauge group remains unknown. The globally consistent forms of the Standard Model gauge group that are both experimentally  and theoretically viable are 
\begin{equation}\label{eqn:GSM}
G_\text{SM}=
\left(SU(3)_C\times SU(2)_W\times U(1)_Y \right)/\Gamma~,
\end{equation}
where $\Gamma=\mathbb{Z}_n\subset \mathbb{Z}_6$ is in the center of the numerator.\footnote{   $\mathbb{Z}_6\subset SU(3)_C\times SU(2)_W\times U(1)_Y $  is the maximal subgroup  that leaves 
all Standard Model fields invariant (see table \ref{tab:StandardModelquantumnumber2} for Standard Model quantum numbers). $\mathbb{Z}_6$   acts   as $z_3\otimes z_2\otimes e^{2\pi i q_Y/6}$, where $z_{3}$ and $z_2$ are the generators of the center of $SU(3)_C$ and $SU(2)_Y$ respectively and $q_Y$ is the hypercharge assignment.}
 Thus, the Standard Model gauge group can take one of four possible forms, labeled by $\Gamma=1,\mathbb{Z}_2,\mathbb{Z}_3,$ or $\mathbb{Z}_6$, one for each     subgroup  of $\mathbb{Z}_6$.
For recent discussions about the global form of Standard Model gauge group see e.g. \cite{Tong:2017oea,Davighi:2019rcd,Buen-Abad:2021fwq,PhysRevLett.132.121601,Reece:2023iqn}

Each global form of the  Standard Model  gauge group  has  a distinct spectrum of line operators~\cite{Tong:2017oea}. For instance, since Wilson lines are 
in linear representations of $G_\text{SM}$, the  Wilson line in the representation $(\mathbf{3},\mathbf{2},\mathbf{1}_{0})$ of $G_\text{SM}$ exists   in the theory with $\Gamma=1$, but is 
 absent for all other choices of the Standard Model gauge group.
 Different global forms also lead to different fractional instanton numbers, and therefore to different periodicity of theta angles (e.g. \cite{Gaiotto:2014kfa,Hsin:2018vcg}) and quantization conditions on the axion couplings \cite{PhysRevLett.132.121601,Reece:2023iqn}.
 
This work shows that the global form of the Standard Model gauge group $G_{\text{SM}}$ can be distinguished by the fractional topological transport coefficient $ \xi$   involving
the  $U(1)$  B-L (0-form) symmetry and the $U(1)$ one-form magnetic symmetry  of the renormalizable Standard Model.\footnote{Henceforth, global symmetry  means the   global symmetry of the
renormalizable Standard Model.}
The fractional response $\xi$ crucially depends on the global form of gauge group $G_{\text{SM}}$, that is on the choice of $\Gamma=\mathbb Z_n$, and on the choice of action of the B-L symmetry, which is not unique due to the electric one-form symmetry of the Standard Model.

In the Standard Model with gauge group $G_{\text{SM}}$, the presence of a background gauge field $A$ for the $U(1)$  B-L symmetry induces a two-form current associated with the 
$U(1)$ one-form magnetic symmetry of the Standard Model given by 
\begin{equation}
    J^{(2)}=\xi\star \frac{dA}{2\pi}\,,
    \end{equation}
   with
 \begin{equation}  
\xi=\frac{n}{3}+\frac{kn}{6}\text{ mod }1~, 
\end{equation}
where 
$n$ can take values $1,2,3$ and $6$ depending on the choice of $G_{\text{SM}}$, $k=0,1,2,\cdots, (6/n)-1$ labels a choice of fractionalization class,  required to uniquely determine the action of B-L on the Standard Model  and $\star$ is the Hodge dual.\footnote{See section 2 for a discussion of   symmetries and fractionalization classes in the Standard Model.} Table \ref{tab:kappa-GSM} contains the  values of $\xi$ for each   choice of  gauge group    $G_{\text{SM}}$ and fractionalization class. 
\begin{table}[h]
    \centering
    \begin{tabular}{|c|c|c|c|c|}
\hline
        $\xi$ &  $\Gamma=1$ & $\Gamma=\mathbb{Z}_2$ & $\Gamma=\mathbb{Z}_3$ & $\Gamma=\mathbb{Z}_6$ \\ \hline
         Integer & \cmark & \cmark & \cmark & \cmark\\
        Integer $+1/2$ & \cmark & \xmark & \cmark & \xmark\\
        Integer $\pm 1/3$ & \cmark & \cmark & \xmark & \xmark\\
        Integer $\pm 1/6$ & \cmark & \xmark & \xmark & \xmark\\ \hline
    \end{tabular}
    \caption{The fractional part of the topological response $\xi$ for the B-L symmetry and magnetic one-form symmetry constrains the global form of the Standard Model gauge group $G_\text{SM}=\left(SU(3)\times SU(2)\times U(1)\right)/\Gamma$. In particular, if $\xi$ is an integer$\pm 1/6$, the Standard Model gauge group is uniquely determined by $\xi$.}
    \label{tab:kappa-GSM}
\end{table}

In   components, the induced charge density and current take  the form
\begin{equation}
\label{inducurre}
    \rho_{0i}=\xi {B_i\over 2\pi},\quad 
    J_{ij}=\xi \epsilon_{ijk}{E^k\over 2\pi}~,
\end{equation}
where $\rho_{0i}=J_{0i}$, and $E^i,B^{i}$ are the background ``electric field" and ``magnetic field" for the $U(1)$ B-L symmetry. This can be contrasted with  the topological response in the Fractional Quantum Hall Effect in 2+1d
\begin{equation}
\label{inducurreFQHE}
    \rho_{0}=\sigma_{xy}  {B_z },\quad 
    J_{x}=\sigma_{xy}{E^y }~,
\end{equation}
where $\sigma_{xy}=\nu/2\pi$ is the fractional Hall conductance for filling fraction $\nu$ and $E$ and $B$ are the background elecromagnetic field. 
Topological transport\footnote{The transport of various generalized symmetries have been studied in e.g. \cite{Das:2023nwl,Hsin:2024aqb}.} coefficients     probe    the   fractionally charged  excitations  of a physical system, such as anyons in the Fractional Quantum Hall Effect. 
In particle physics, quarks are the    fractionally charged excitations under the B-L global symmetry of the Standard Model.

The fractional topological response $\xi$ in the Standard Model can be detected by a local correlation function. It is captured by  a contact term in the  following current two-point function in Euclidean space
\begin{equation}\label{eqn:contacterm}
    \langle J^{(1)}_\mu(x) J^{(2)}_{\nu\lambda}(0)\rangle =i\frac{\xi}{2\pi}\epsilon_{\mu\nu\lambda\rho}\partial^\rho \delta^{(4)}(x)+\cdots~,
\end{equation}
where    $J^{(1)}$ is the current of the $U(1)$ B-L symmetry,  $J^{(2)}$ is the current of the $U(1)$ one-form magnetic symmetry and $\cdots$ are  non-contact terms. Since the contact term can be shifted by an integer $\xi\rightarrow\xi+1$ by adding a well-defined local counterterm   corresponding to stacking a symmetry protected topological (SPT) phase for  $U(1)$ 0-form and $U(1)$ one-form symmetries (see  e.g. \cite{Jian_2021}),
 it is the fractional part of $\xi$ that is   physical and scheme independent.
This is analogous to the contact term in the current two-point function that captures  Hall conductivity in 2+1d \cite{Closset:2012vp}. 

In summary, the work shows that the global form of the Standard Model gauge group $G_{\text{SM}}$ can be distinguished by the fractional topological transport coefficient $ \xi$   involving
the  $U(1)$  B-L   symmetry and the $U(1)$ one-form magnetic symmetry  of the Standard Model. It can be detected as an induced current in the presence of a background gauge field for the B-L symmetry (see (\ref{inducurre}) or in a correlation function of conserved currents. This begs the question whether    the topological transport in the Standard Model introduced in the work is measurable in any foreseeable  experiment. This is  an interesting and important question deserving of attention. A preliminary  speculation is  that the induced current could be detected around a macroscopic source for  B-L,  perhaps near a neutron star or   closer to home, on earth. This may require considering a very weakly gauged B-L symmetry, a scenario very well studied in the literature and consistent with   experimental data.   The detectability of $\xi$ is an important open question for the future.

The plan of the rest of the paper is as follows. Section \ref{sec:symm} discusses the  0-form and one-form  symmetries of the Standard Model with gauge group $G_{\text{SM}}$ and briefly discuss  the physics  of symmetry fractionalization. Section \ref{sec:toptransport} introduces the notion of topological transport in 3+1d involving a $U(1)$ 0-form symmetry and a $U(1)$ one-form symmetry. Section \ref{sec:frac} 
computes the 
the   fractional topological transport coefficient $ \xi$ in the Standard Model for each choice of gauge group $G_{\text{SM}}$ and fractionalization class.

\section{ Standard Model Symmetries   and Symmetry Fractionalization}
\label{sec:symm}

The section will start with a brief discussion of how gauge redundancies and global symmetries are realized on matter fields and gauge-invariant operators in quantum field theory. In gauge theories  the normalization of the gauge charges  should be such that   matter fields carry a linear representation  of the gauge group,  and not   a    projective representation.
This is the normalization used in table \ref{tab:StandardModelquantumnumber2}, where all fields have properly quantized gauge charge.

Gauge-invariant local operators must furnish a    linear representation of the global 0-form symmetry of the quantum field theory. This is in stark contrast to the global symmetry transformations of elementary fields charged under gauge redundancies. Gauge-variant fields  can transform in a projective representations of the   {\it global} 0-form symmetry. This observation is central in this work. I proceed to discussing relevant symmetries of the Standard Model and how they depend of the choice of gauge group $G_{\text{SM}}$.

\paragraph{Zero-form symmetry:}
The classical  renormalizable Standard model admits the action of a $U(1)$ baryon symmetry $B$ and a $U(1)$ lepton symmetry $L_i$ for each generation.\footnote{There are Standard Model higher dimension operators that break these symmetries.  $B$ and $L_i$ are accidental symmetries of the renormalizable Standard Model. The current measurements of the neutrino mass matrix are consistent with an unbroken B-L symmetry.} The ABJ anomalies of the Standard Model imply that  
B-L    survives as a $U(1)$ global 0-form symmetry after quantization. In terms of the quantum numbers in table \ref{tab:StandardModelquantumnumber2}, the conserved charge of the B-L symmetry is
\begin{equation}\label{eqn:B-Lcharge}
\text{Global 0-form $U(1)$ symmetry}:\quad     Q=\frac{1}{3} B-\sum_i L_i~.
\end{equation}
In this  normalization, the   gauge invariant operators  with minimal charge in  the Standard Model (such as $L_i^\dag H$ or baryon operator) carry charge $Q=1$.
Instead, the  quarks $Q_i,\bar u_i,\bar d_i$ transform projectively under the    global  0-form $U(1)$ B-L symmetry, and carry fractional charges under $U(1)$ B-L.\footnote{
The fractional charge here is not to be confused with the often-stated fractional gauge charges of the Standard Model matter fields (see e.g. \cite{annurev:/content/journals/10.1146/annurev-nucl-121908-122035,Langacker:2011db}), which are not properly quantized.}   

\begin{table}[t]
    \centering
    \begin{tabular}{|c|ccc|cc|c|}
\hline
         & $Q_i$ & $\bar u_i$ & $\bar d_i$ & $L_i$ & $\bar e_i$ & $H$ \\ \hline
      $SU(3)_C$   & $\mathbf{3}$ & $\overline{\mathbf{3}}$ & $\overline{\mathbf{3}}$ &  & &\\
      $SU(2)_W$ & $\mathbf{2}$ & & & $\mathbf{2}$ & & $\mathbf{2}$\\
      $U(1)_Y$ & $+1$ & $-4$ & $+2$ & $-3$ & $+6$ & $-3$\\ \hline 
      $U(1)_B$ & $+1$ & $-1$ & $-1$ & & & \\
      $U(1)_{L_i}$ & & & & $+1$ & $-1$ & \\ \hline
     Global $Q$ &$1/3$ & $-1/3$ & $-1/3$ & $-1$ & $+1$ &  \\ \hline
    \end{tabular}
    \caption{Quantum numbers of  quarks, leptons and Higgs in the Standard Model. The index $i$ labels the three generations, $i=e,\mu,\tau$. The $U(1)_Y$ charges are normalized such that the minimal charge is one. All fermion fields are left-handed, with the barred fields denoting the charge conjugate of right handed fields. $U(1)_B$ and $U(1)_{L_i}$ are the classical baryon and lepton number symmetries of the   Standard Model. The last row is the $U(1)$ B-L global symmetry charge $Q$ in (\ref{eqn:B-Lcharge}), which survives quantization of the renormalizable Standard Model.}
\label{tab:StandardModelquantumnumber2}
\end{table}

The quarks in  the Standard Model behave analogously  to the quasiparticles in the  fractional quantum Hall effect. In the fractional quantum Hall state described by emergent  Chern-Simons gauge fields, the quasiparticles transform in a linear representation of the Chern-Simons gauge group and   a projective representation of the global $U(1)$ symmetry.\footnote{The particles transforming in projective representation of the Chern-Simons gauge group are confined and they do not appear in the fractional quantum Hall effect.} Indeed, quarks transform in a linear representation of $G_{\text{SM}}$ and   in a projective representation of  the  $U(1)$ B-L symmetry.

\paragraph{One-form symmetry:} A one-form symmetry is a global symmetry   that acts      on line operators instead of local operators, and give rise to   Ward  identities and selection rules \cite{Gaiotto:2014kfa}.
Line operators transform in a linear representation of the one-form symmetry group just as local gauge invariant operators transform in a linear representation of the 0-form symmetry.

The Standard Model with $G_\text{SM}=\left(SU(3)\times SU(2)\times U(1)_Y\right)/\Gamma$ gauge group has both electric and magnetic one-form symmetry:
\begin{equation}
\text{Global one-form symmetry}:\quad
    U(1)\times (\mathbb{Z}_6/\Gamma)~,
\end{equation}
where $U(1)$ is the magnetic one-form symmetry from the hypercharge gauge group $U(1)_Y$, and acts on the 't Hooft lines, 
and $(\mathbb{Z}_6/\Gamma)$ is the electric one-form symmetry, and acts on the Wilson lines.\footnote{The discrete dual magnetic symmetry $\Gamma\cong \text{Hom}(\Gamma,U(1))$ due to the quotient of the gauge group $G_\text{SM}$ is a subgroup of the magnetic $U(1)$ one-form symmetry.} 

The continuous $U(1)$ magnetic one-form symmetry is generated by a two-form current \cite{Gaiotto:2014kfa}
\begin{equation}
    J^{(2)}=\star n \frac{da_Y}{2\pi}~,
    \label{eqn:currenttt}
\end{equation}
where $a_Y$ is the gauge field for the hypercharge $U(1)_Y$.  The current is conserved, obeying $d\star J^{(2)}=0$,  by virtue of the Bianchi identity. Due to the quotient by $\mathbb Z_n$ in $G_{\text{SM}}$ (\ref{eqn:GSM}), the hypercharge gauge field $a_Y$ can have fractional magnetic charge given by a multiple of $1/n$, and thus the normalization factor $n$ in $J^{(2)}$  ensures that the global one-form symmetry charge $\oint \star J^{(2)}$ of the minimally charged line operator is an integer with minimal charge  $1$.

\paragraph{Fractionalization classes:} The following is a brief review of the concept of symmetry fractionalization (see e.g. \cite{Barkeshli:2014cna,Teo_2015,Tarantino_2016,Benini:2018reh,Brennan:2022tyl}) and its consequences for the Standard Model. In the presence of a  higher-form symmetry, a lower-form symmetry requires additional data to completely specify its  action on the physical system. This discrete additional data is known as  a  fractionalization class \cite{Barkeshli:2014cna,Benini:2018reh}.

The action of a  $q$-form symmetry   is implemented by a network of  topological operators of codimension $(q+1)$ in spacetime \cite{Gaiotto:2014kfa}. In order to fully specify the symmetry action  one  must also  specify their junctions where multiple symmetry operators meet. Such junctions, which are of higher codimension $(q'+1)>(q+1)$, can be enriched by the topological    operators of a  $q'$-form symmetry with $q'>q$. Different decorations of the junction using $q'$-form symmetry defects  describe different fractionalization classes, and they give rise to different actions of the $q$-form symmetry. Therefore,  to fully specify a $q$-form symmetry, one needs to choose a fractionalization class for all higher $q'>q$-form symmetries.
Without  specifying the fractionalization class, the $q$-form symmetry remains ambiguous, and can,  for example,    result in ambiguous 't Hooft anomalies for the $q$-form symmetry (see e.g. \cite{Barkeshli:2019vtb,Hsin:2019fhf, Barkeshli:2021ypb}). Thus it is important to specify the fractionalization classes when discussing theories with higher form symmetries, such as   the Standard Model. A fractionalization class admits also a description in terms of     correlated background gauge fields for the $q$ and $q'$-form symmetries (see below), as background fields for global symmetries    provide an an alternative  description of the network of topological operators that implement the symmetry. 

Since the Standard Model has both 0-form and one-form symmetries, the 0-form symmetry requires a choice of fractionalization class. This can be physically interpreted as a choice of    how the B-L symmetry acts on microscopic heavy probe particle (see e.g. \cite{Barkeshli:2014cna}). 
Here the focus is on the fractionalization classes involving the electric one-form symmetry $\mathbb{Z}_6/\Gamma$ of the Standard Model. Given the choice of gauge group $G_{\text{SM}}$,     determined by a discrete group $\Gamma=\mathbb Z_n$, these fractionalization classes are classified by
\cite{Barkeshli:2014cna,Benini:2018reh}
\begin{equation}
k\in H^2(BU(1), \mathbb{Z}_6/\Gamma)\cong \mathbb{Z}_6/\Gamma\cong \mathbb{Z}_{6/n}~.
\end{equation}
Recall that  $n=1,2,3$ or  $6$. The classes are labeled by an integer $k=0,1,\cdots {6\over n}-1$.
In terms of   background gauge fields, a choice of fractionalization class  relates the background two-form gauge field $B^e$ for the electric one-form symmetry $\mathbb{Z}_{6/n}$, 
with the background gauge field $A$ for the 0-form B-L symmetry \cite{Benini:2018reh}
\begin{equation}
    B^e=\frac{k}{(6/n)}dA~,
\end{equation}
with the    normalization $\oint B^e\in \frac{2\pi}{(6/n)}\mathbb{Z}$.

\section{Topological Transport in 3+1d}
\label{sec:toptransport}

This work considers the topological transport for a $U(1)$ 0-form symmetry and a $U(1)$ one-form symmetry in 3+1d. The topological response can be described by  a topological effective action for the $U(1)$ 0-form symmetry and the $U(1)$ one-form symmetry. Denoting the background gauge field for the one-form symmetry by $B$ and the background gauge field for the 0-form symmetry by $A$, the topological effective action is
\begin{equation}
    ``\quad \frac{\xi}{2\pi}\int BdA\quad \text{''}~,
    \label{eq:transp4}
\end{equation}
where the quotation mark is used to denote that the effective action is only properly quantized for integer values of $\xi$. The more precise definition of $\xi$ is via the induced currents below.

The fractional part of $\xi$ is scheme independent, as $\xi$ can be shifted by an arbitrary integer by adding a well-defined local counterterm. The topological transport coefficient $\xi$  is a contact term in the current two-point function  for their one-form and two-form conserved currents $J^{(1)},J^{(2)}$.
Indeed, by taking the functional derivative with respect to the background fields $A,B$ results in the  contact term (\ref{eqn:contacterm})
\begin{equation} 
    \langle J^{(1)}_\mu(x) J^{(2)}_{\nu\lambda}(0)\rangle =i\frac{\xi}{2\pi}\epsilon_{\mu\nu\lambda\rho}\partial^\rho \delta^{(4)}(x)+\cdots~\,.
\end{equation}
Also, taking functional derivatives gives the transport equations
\begin{equation}
  J^{(2)}=\xi\star \frac{dA}{2\pi},\qquad J^{(1)}=\xi \star \frac{dB}{2\pi}~.
  \label{eqn:toptrans}
\end{equation}
 The fractional part of $\xi$ can depend on the choice of a fractionalization class.

Another way to understand the topological transport is the following: from the effective action, one finds that the Poincar\'e dual of the degree-3 integer class $dB/2\pi$, which is a 1-cycle in spacetime, carries fractional charge $\xi$ of the 0-form symmetry. 
Similarly, the Poincar\'e dual of the degree-2 integer class $dA/2\pi$, which is a 2-cycle in spacetime, carries fractional charge $\xi$ of the one-form symmetry.\footnote{
An example of such fractional one-form symmetry charge is the vortex string with $\pi$ flux in $\mathbb{Z}_2$ gauge theory that arises as a Higgs phase of the $U(1)$ gauge theory. The theory has $U(1)$ magnetic one-form symmetry, and since the vortex string has $\pi$ flux, it carries half-integer one-form charge \cite{Hsin:2019fhf}.
} This corresponds to a Dirac string that attaches to an improperly quantized 't Hooft line.
These interpretations can be generalized to the case of discrete symmetries.

The topological transport action (\ref{eq:transp4}) in 3+1d  relevant for the Standard Model has a counterpart in  the  fractional quantum Hall effect in 2+1d. In that context the topological action is  
\begin{equation}
    ``\quad \frac{\nu}{4\pi}\int AdA\quad \text{''}~,
    \label{eq:FQHE}
\end{equation}
where $\sigma_{xy}=\nu/2\pi$ is the celebrated fractional Hall conductance and $A$ is the background gauge field for the $U(1)$ 0-form symmetry. $\nu$ is the filling fraction.  Again,   the quotation mark is used to denote that the effective action is only properly quantized for integer values of $\nu$. The well-defined   topological response Hall current is given in equation (\ref{inducurreFQHE}).

\section{Fractionalization in Standard Model Detects Gauge Group}
\label{sec:frac}

This section computes the   topological transport coefficient $\xi$ 
involving
the  $U(1)$  B-L  symmetry and the $U(1)$ one-form magnetic symmetry  of the  Standard Model. As will be shown below, $\xi$ depends on the choice of gauge group $G_{\text{SM}}$ and  of fractionalization class. 

\subsection{Selection rule of fractional B-L charge}

A key observation follows from looking at the gauge and B-L quantum numbers of the Standard Model fields   in table  \ref{tab:StandardModelquantumnumber2}.   The quantum numbers  obey a  $\mathbb{Z}_3$ selection rule relating the  $U(1)_Y$ gauge charge with the fractional part of the B-L symmetry charge $Q$:
\begin{equation}
    Q=\frac{1}{3}q_{U(1)_Y}\text{ mod }1~.
\end{equation}
This selection rule is realized nontrivially on the quarks.

This selection rule implies that  the 0-form symmetry structure of the classical Lagrangian of Standard Model is
\begin{equation}\label{eqn:quotientsymmetrySM}
    \frac{SU(3)_C\times SU(2)_W\times U(1)_Y\times \tilde U(1)_\text{B-L}}{\Gamma\times \mathbb{Z}_3}=\frac{G_\text{SM}\times \tilde U(1)_\text{B-L}}{\mathbb{Z}_3}~,
\end{equation}
where $U(1)_\text{B-L}=\tilde U(1)_\text{B-L}/\mathbb{Z}_3$ is the B-L symmetry. The $\mathbb{Z}_3$ quotient in (\ref{eqn:quotientsymmetrySM}) identifies the $\mathbb{Z}_3$ subgroups of $U(1)_Y$ and $\tilde U(1)_\text{B-L}$. 

This observation has far reaching consequences for the  topological transport coefficient $\xi$ of the Standard Model.

\subsection{Twisted gauge fields}

The gauge fields for the symmetry structure (\ref{eqn:quotientsymmetrySM}) is discussed below, following e.g.    \cite{Benini:2017dus,Cheng:2022nji}.
The gauge fields for $G_\text{SM}/\mathbb{Z}_3$ are twisted gauge fields for $(SU(3)_C\times SU(2)_W)/\Gamma$ and $U(1)_Y/\Gamma'$, where $\Gamma=\mathbb{Z}_n$ and $\Gamma'=\mathbb{Z}_{\text{lcm}(n,3)}$. This implies that gauge fields are correlated and  that 
the $U(1)_Y$ gauge field $a_Y$ has fractional flux 
\begin{align}
\label{eq:twistedf}
    \frac{da_Y}{2\pi}&=\frac{1}{\text{lcm}(n,3)}\left(\frac{3}{\gcd(n,3)} w_2^{(n)}+\frac{n}{\gcd(n,3)}\frac{dA}{2\pi}\right)\cr 
    &=\frac{1}{n}w_2^{(n)}+\frac{1}{3}\frac{dA}{2\pi}~,
\end{align}
where $w_2^{(n)}$ is the $\Gamma=\mathbb{Z}_n$ 2-cocycle obstruction class to lifting an $(SU(3)_C\times SU(2)_W)/\Gamma$ gauge field to an $SU(3)_C\times SU(2)_W$ gauge field. $A$ is the background gauge field for the B-L symmetry. Therefore, 
turning on a background gauge field for  B-L  induces fractional flux for $U(1)_Y$, which would otherwise be integer quantized in the absence of $A$.

\subsection{Topological response with zero fractionalization class}

In the absence of a background gauge field $A$ for the B-L 0-form symmetry,   the current of the properly quantized magnetic $U(1)$ one-form symmetry in the Standard Model with gauge group $G_{\text{SM}}$ is (\ref{eqn:currenttt})
\begin{equation}
    J^{(2)}=n\star \frac{da_Y}{2\pi}~\,,
\end{equation}
such that $\oint \star J^{(2)}$ is an integer  with minimal charge $1$. Recall that $n$ can be $1,2,3$ or $4$ depending on choice of gauge group  $G_{\text{SM}}$.

Turning on a gauge field  $A$ for B-L induces a nontrivial current for the one-form global magnetic symmetry by virtue of (\ref{eq:twistedf})
\begin{equation}
    J^{(2)}=\xi \star \frac{dA}{2\pi}~\,.
\end{equation}
This implies that 
 the Standard Model gauge group with $\Gamma=\mathbb Z_n$ has the following  topological response $\xi$  
\begin{equation}
    \xi=\frac{n}{3}\text{ mod }1~.
\end{equation}
This follows from 
 equation (\ref{eqn:toptrans}).
 
\subsection{Topological response for general fractionalization class}

The topological response $\xi$ for general choice of fractionalization class is computed below. Another key element in this analysis is the mixed anomaly between the electric and magnetic one-form symmetries of the Standard Model.

\paragraph{Fractionalization class}

The fractionalization classes can be changed by turning  on a   background gauge field 
\begin{equation}\label{eqn:fracBe}
    B^e=\frac{k}{(6/n)}dA~,
\end{equation}
 for the $\mathbb{Z}_{6/n}$ electric one-form symmetry of the Standard Model with gauge group $G_{\text{SM}}$ for choice $\Gamma=\mathbb Z_n$.  $k=0,1,\cdots, (6/n)-1$ is an integer that labels the fractionalization class \cite{Benini:2018reh}.

The physical meaning of  (\ref{eqn:fracBe})  is as follows.
Consider the topological junction of $U(1)$ B-L 0-form symmetry generators $[\theta_1],[\theta_2],[\theta_1+\theta_2]$ where $\theta_i$ are angular parameters that label the $U(1)$ transformations, and $[\cdot]$ is the restriction to $[0,2\pi)$. The relation (\ref{eqn:fracBe}) implies that at the (codimension-two) junction there is an additional generator for the electric one-form symmetry with group  element
\begin{equation}
    k\frac{\theta_1+\theta_2-[\theta_1+\theta_2]}{2\pi}\text{ mod }(6/n)\in\mathbb{Z}_{6/n}~.
\end{equation}

\paragraph{Additional topological transport}

The electric and magnetic one-form symmetries of the  Standard Model have a mixed 't Hooft anomaly. This is due to the mixed anomalies between the electric and magnetic one-form symmetries of the $U(1)_Y$ gauge field, as discussed in \cite{Gaiotto:2014kfa,Hsin:2019fhf}.

The mixed 't Hooft anomaly can be described by a bulk topological term
\begin{equation}
\label{eqn:anom}
    \frac{1}{2\pi} B^e dB~,
\end{equation}
where $B$ is the background for the $U(1)$ magnetic one-form symmetry.
For the fractionalization class $k$, one substitutes (\ref{eqn:fracBe}) in (\ref{eqn:anom}), which produces a total derivative term
\begin{equation}
    \frac{\Delta \xi}{2\pi} dAdB,
    \end{equation}
    with
 \begin{equation} \Delta \xi=\frac{k}{(6/n)}=\frac{kn}{6}\text{ mod }1~.
\end{equation}
The total derivative term in the bulk can be canceled by a boundary term, leading to a shift of the fractional topological transport $\xi\rightarrow \xi+\Delta \xi$. 

Therefore, the  fractional topological transport $\xi$ for a choice of gauge group (choice of $\Gamma=\mathbb Z_n$) and fractionalization class $k$ is 
\begin{equation}
\xi=\frac{n}{3}+\frac{kn}{6}\text{ mod }1~, 
\end{equation}
with $k=0,1,\cdots, (6/n )-1.$

As discussed in e.g. \cite{Tong:2017oea}, the Standard Model can have continuous or discrete topological theta terms. The continuous theta terms are total derivatives. The discrete theta term arise from the quotient (gauging) of $SU(3)_C\times SU(2)_W\times U(1)_Y$ by $\Gamma$, and it can take torsional values $p\in\mathbb{Z}_{2n}$ for $\Gamma=\mathbb{Z}_n$ \cite{Hsin:2018vcg}:
\begin{equation}
    2\pi \frac{p}{2n}\int {\cal P}(w_2^{(n)})~,
\end{equation}
where ${\cal P}$ is the Pontryagin square.
Importantly, none  of these topological terms contribute to the fractional topological transport coefficient $\xi$, which is a nontrivial contact term. One can also check it explicitly using the above topological action.
 
Below is the complete list of  possible fractional values of $\xi$,  and what choice of gauge group $\Gamma=\mathbb{Z}_n\subset \mathbb{Z}_6$ and fractionalization class $k\in k\in\mathbb{Z}_{6/n}$ can give rise to that $\xi$:
\begin{itemize}
    \item If $\xi\in \mathbb{Z}+1/6$, then $\Gamma=1,k=5$.
      \item If $\xi\in\mathbb{Z}+5/6$, then $\Gamma=1,k=3$.
    \item If $\xi\in\mathbb{Z}+1/3$, then either $\Gamma=1,k=0$ or $\Gamma=\mathbb{Z}_2,k=2$.
    \item If $\xi\in\mathbb{Z}+2/3$, then either $\Gamma=1,k=2$ or $\Gamma=\mathbb{Z}_2,k=0$.
    \item If $\xi\in \mathbb{Z}+1/2$, then  $\Gamma=1,k=1$ or $\Gamma=\mathbb{Z}_3,k=1$.
    \item If $\xi\in\mathbb{Z}$,   $\xi$ has no   fractional part, then  $\Gamma=1, k=4$, or  $\Gamma=\mathbb Z_2, k=1$ or $\Gamma=\mathbb{Z}_3,k=0$, or $\Gamma=\mathbb{Z}_6$.
\end{itemize}
The relation between the fractional part of $\xi$ and the Standard Model gauge group $G_{\text{SM}}$ is summarized in table \ref{tab:kappa-GSM}.

\section*{Acknowledgments}

I thank Matthew Buican and Shota Komatsu for discussion about symmetry in Standard Model. 
P.-S.H. is supported by Department of Mathematics, King's College London. I thank CERN for invited presentation related to this work, during which part of the work is completed.
Research of G. at Perimeter Institute is supported in part by the Government of Canada through the Department of Innovation, Science and Economic Development Canada and by the Province of Ontario through the Ministry of Colleges and Universities.

\vfill\eject

\appendix

\bibliographystyle{utphys}
\bibliography{biblio}

\end{document}